\documentclass{article}
\usepackage{ijcai25}

\usepackage{times}
\usepackage{soul}
\usepackage{url}
\usepackage[hidelinks]{hyperref}
\usepackage[utf8]{inputenc}
\usepackage[small]{caption}
\usepackage{graphicx}
\usepackage{amsmath}
\usepackage{amsthm}
\usepackage{booktabs}
\usepackage{algorithm}
\usepackage{algorithmic}
\usepackage[switch]{lineno}

\title{Urban Emergency Rescue Based on Multi-Agent Collaborative Learning: Coordination Between Fire Engines and Traffic Lights }

\author{
Weichao Chen$^1$ \and
Xiaoyi Yu$^1$ \and
Longbo Shang$^1$ \and
Jiange Xi$^2$ \and
Bo Jin$^1$\and
Shengjie Zhao$^1$\\
\affiliations
$^1$School of Computer Science and Technology, Tongji University\\
$^2$School of Computer Science, Fudan University
}

\begin{document}

\maketitle

\begin{abstract}
Nowadays, traffic management in urban areas is one of the major economic problems. In particular, when faced with emergency situations like firefighting, timely and efficient traffic dispatching is crucial. Intelligent coordination between multiple departments is essential to realize efficient emergency rescue. In this demo, we present a framework that integrates techniques for collaborative learning methods into the well-known Unity Engine simulator, and thus these techniques can be evaluated in realistic settings. In particular, the framework allows flexible settings such as the number and type of collaborative agents, learning strategies, reward functions, and constraint conditions in practice. The framework is evaluated for an emergency rescue scenario, which could be used  as a simulation tool for urban emergency departments.
\end{abstract}

\section{Introduction}

 Urban traffic management is a critical research area in the domain of Smart City, which needs strategies to mitigate congestion, reduce idle times, and promote smoother traffic flow in the city \cite{b1,b2,b3}. In particular, when faced with emergency situations like firefighting, timely and efficient traffic dispatching is crucial. Urban emergency rescue is one of the key activities that reflects the emergency response capabilities of urban management departments \cite{b4}. It requires coordinated decision-making across multiple departments to swiftly aid individuals while minimizing the impact on others. In such critical tasks, traffic management must seamlessly integrate real-time urban operational data to ensure efficient traffic flow. This integration of real-time data is essential for optimizing emergency response routes, prioritizing emergency vehicles' passage, and dynamically adjusting traffic signals to facilitate the rapid movement of rescue teams. 

In the context of urban traffic management, the dynamic and complex nature of traffic scenarios demands innovative solutions that go beyond traditional paradigms. Conventional traffic management approaches \cite{b5,b6,b7,b8} predominantly focus on the independent decision-making of single-type intelligent agents, ignoring the potential benefits that collaborative mechanisms among multiple types of agents could offer. Although single-type intelligent agents can make decisions based on local information and objectives, the integration of multiple types of agents with diverse capabilities and perspectives can lead to more robust and adaptive traffic management strategies. By harnessing the power of multi-type collaborative mechanisms, urban traffic management systems could achieve enhanced efficiency, resilience, and responsiveness in the face of emergencies such as firefighting operations. 

Multi-agent reinforcement learning (MARL) has shown its superior performance in the field of autonomous unmanned systems, for it promotes agents' information sharing and collective strategy optimization, thereby enhancing the overall efficiency of the learning framework \cite{b10}. Additionally, MARL's adaptability to dynamic environments allows agents to adjust their strategies in response to the feedback from their peers, providing robustness in performance. In \cite{b11}, the authors proposed a novel multi-agent reinforcement learning model, which optimizes the route length and the vehicle’s arrival time simultaneously. The work of 3M-RL \cite{b12} used an Actor-Critic neural network to obtain the routing policy. In that scenario, each UAV makes decisions based on their local observations, to reach their destinations as soon as possible while avoiding collisions. 

In this demo, we present a collaborative decision-making system composed of multiple types of intelligent agents (e.g., fire engines, ambulances, and traffic lights), in which we employ the MARL algorithm to minimize rescue losses. Specifically, in our investigated system, we carefully design the reward function of the MARL to accelerate the convergence speed while ensuring the actions these agents take do not violate actual rules and physical characteristics. We conduct preset route planning for fire engines and uniformly configure modules such as vehicle control, navigation, and modeling visualization in the Unity Engine simulator. To further promote the related research on traffic management of Smart City in the community, we make this simulation environment an open-source project. The code is available at \emph{https://github.com/Yxy54321/TrafficSim-MARL}.

\section{System Model and Architecture}
To implement multi-agent collaborative learning in a smart city, we deploy elements such as buildings, vehicles, traffic lights, etc., based on the Unity Engine. The vehicles in the simulated system are composed of three parts: ordinary vehicles, noise vehicles, and special vehicles (e.g., fire engines). In the following, we first present the environment setting in the Unity Engine, and then propose a novel multi-agent collaborative learning algorithm that can be deployed on fire engines and traffic lights. Finally, several constraint conditions are considered to make the simulation environment more compatible with the actual scenario. Fig. 1 shows the procedure of interaction through HTTP Server between multiple agents and the simulated environment.

\subsection{Smart City Based on the Unity Engine}

\paragraph{Ordinary Vehicles}
Vehicles with fixed starting points and destinations, exhibiting regular movement patterns. Their starting points and destinations are automatically determined by the system at program initialization, and their actual routes are generated by the navigation system's path planning algorithms. These residents follow system-generated navigation paths with a uniform distribution of trajectories, showing no distinct clustering characteristics.

\paragraph{Noise Vehicles}
Additional vehicles artificially introduced to simulate realistic traffic conditions, as ordinary residents alone create overly balanced traffic flows. These vehicles are strategically placed to generate controlled congestion at specific intersections and road sections, creating time-varying traffic patterns that better reflect real-world situations. The system utilizes the navigation system's path planning algorithms for navigation, with designated start and end points to simulate these dynamic traffic state changes.

\paragraph{Special Vehicles}
Vehicles that are used for providing emergency services (e.g., firefighting, medical service), whose route planning is controlled by the proposed reinforcement learning algorithms. The system implements a separate Unity Engine thread that communicates with an external Python program through APIs, allowing intelligent path planning and adaptive control based on the learned strategies. 

\begin{figure}[htbp]
\centerline{\includegraphics[width=\linewidth]{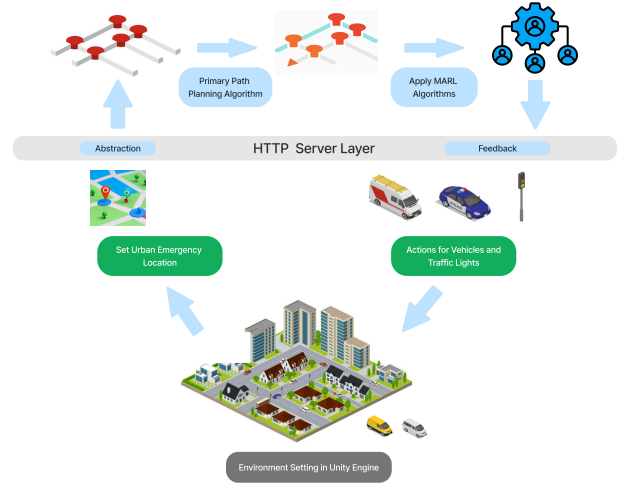}}
\caption{System architecture.}
\label{fig}
\end{figure}

\subsection{Multi-Agent Collaborated Learning Between Fire
Engines and Traffic Lights}

\paragraph{MARL}
We employ the idea of QMIX\cite{b14} algorithm to achieve multi-agent collaborative learning (MARL) between fire engines and traffic lights. QMIX is designed for multi-agent reinforcement learning, aimed at optimizing global rewards through a mechanism of centralized training and decentralized execution, particularly suitable for complex tasks that require collaboration among multiple agents.

\paragraph{Reward Function Design}

The design of the reward function is crucial in multi-agent reinforcement learning. A reasonable reward function can not only accelerate the convergence of the model but also guide the agents to learn the optimal strategy that aligns with the task objectives. For the tasks of fire truck path planning and traffic signal cooperation control, we carefully design the reward function as follows

\begin{enumerate}
    \item \textbf{Goal Achievement}: When the fire engines successfully reaches the target location, a reward of 100 is given. This reward clarifies the mission goal and encourages the agents to work towards the target location.
    
    \item \textbf{Collision Avoidance}: If a fire truck collides with other vehicles, deduct 50. This design aims to reduce high-risk decision-making and encourage agents to learn obstacle avoidance behavior.
    
    \item \textbf{Step Penalty}: Every time step, a deduction of 0.1 is applied. The purpose of this penalty is to encourage the agent to complete the task as quickly as possible, avoiding time waste caused by inefficient exploration.
    
    \item \textbf{Approach Reward}: If the fire truck is closer to the target position compared to the previous time step, a reward is provided based on the proximity distance \(\Delta d\).
    \begin{equation}
    r = \min(\alpha \cdot \Delta d, 3).
    \end{equation}
    \(\alpha\) is the reward coefficient, which is used to control the intensity of the reward.The design ensures that the single-step reward value does not exceed 3 to avoid negative impacts on the training process due to excessively large reward values.
\end{enumerate}

The designed combinatorial reward mechanism encourages agents to achieve task objectives while effectively avoiding potential risk behaviors through the balance of positive and negative feedback. By stepwise rewards, agents can optimize each step's actions progressively while completing the task, thereby learning the globally optimal strategy.

\section{Simulation Settings and Results}
The simulation system consists of four key modules: Vehicle Control, Navigation, Modeling, and Server, which collectively enable the simulation and control of vehicle behavior within the Unity Engine. The vehicle control module implements the atomic vehicle operations, serving as the fundamental control layer for vehicle behavior. It employs physical engine ray detection for environmental sensing and collision awareness, while converting navigation commands into basic physical actions, i.e., straight-line movement and turning maneuvers. Besides control implementation, it continuously collects vehicle observations through ray detection, including position information and surrounding environment data, and transmits these observations to the server module. The navigation module abstracts real-world road conditions into a graph structure to implement efficient path planning algorithms. The navigation strategies vary according to different types of vehicles. The modeling module realizes 3D visualization and simulation of a Smart City, including the modeling of buildings, roads, and traffic lights at intersections. Based on the Unity's physics engine, it facilitates vehicles' interactions through the navigation module, obstacle avoidance through ray detection, and provides the effects of special vehicles upon collisions. The server module provides encapsulated APIs for external Python programs. The interfaces enable the deep network to interact with the environment, reflect actions, and receive feedback in the form of rewards for back propagation and network updates. The module operates on a separate thread to make sure asynchronous communication, and thus avoids interference with Unity’s physics and rendering processes. The training cycle repeats until the termination condition is met or the maximum step is reached, after which the program samples the data for training and sends a reset request to continue.

To demonstrate the effectiveness of the proposed QMIX training framework with the specified reward function, we show the rewards comparison results with the IQL strategy in Fig. 2. We can observe that QMIX obtains a much higher reward compared to IQL-based method since it further utilized the global information exchanged between the fire engines and traffic lights. In addition, as more agents (2 Fire Engines and 16 Traffic Lights) are involved in this process, QMIX could generate better decisions by obtaining more valuable information.

\begin{figure}[htbp]
\centerline{\includegraphics[width=\linewidth]{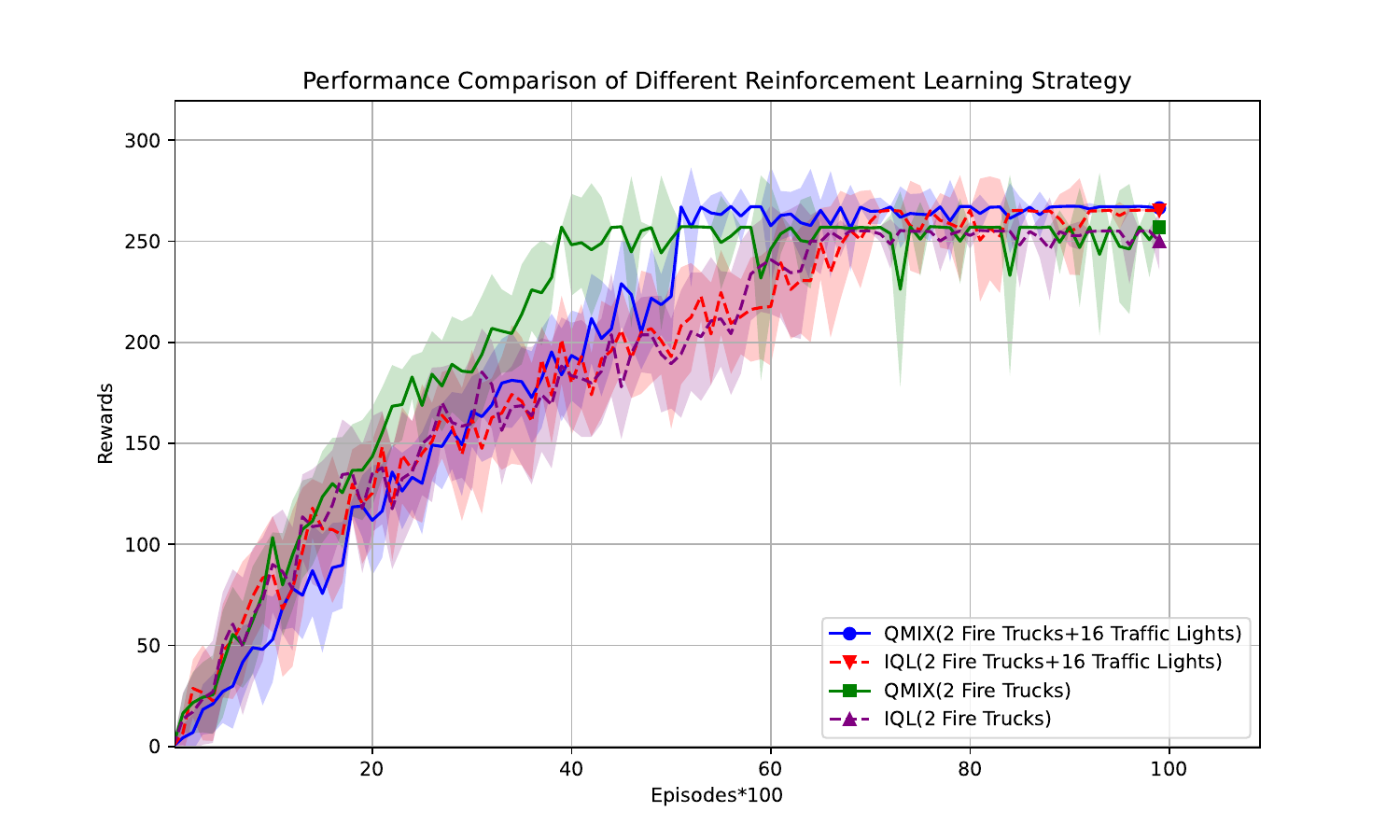}}
\caption{Rewards comparison between different strategies.}
\label{fig}
\end{figure}

At the beginning of the simulation, we first set the location of the fire outbreak, the positions of the fire engines, and the distribution of traffic lights. Then, we select a collaborative reinforcement learning method for all agents and calculate the current reward values and the elapsed time for each agent in real-time. Finally, when the fire engines reach the designated fire outbreak points, the simulation process ends. An illustration of such simulation process can be seen in Fig. 3.

\begin{figure}[htbp]
\centerline{\includegraphics[width=\linewidth]{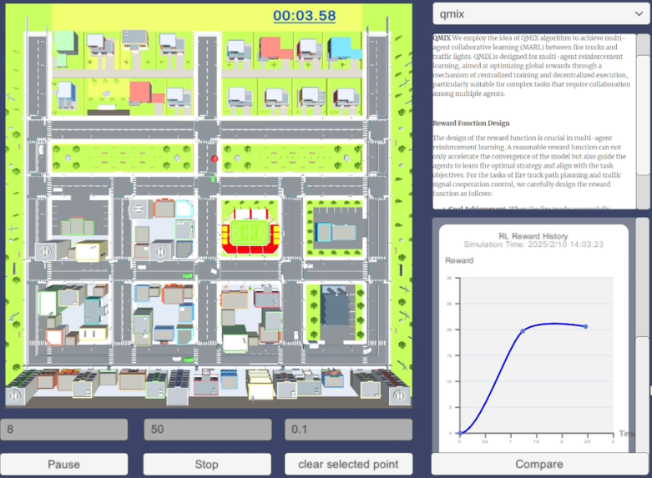}}
\caption{Screenshot of the simulator in Unity Engine.}
\label{fig}
\end{figure}

\section{Conclusion}
This demonstration investigated a collaborative decision-making system composed of multiple types of intelligent agents, and proposed to apply the MARL algorithm in the scenario of urban emergency rescue. The fire engines and traffic lights are jointly optimized to minimize the overall rescue losses. By carefully designing the reward function according to the practical constraint condition, each agent would be encouraged to achieve task objectives while avoiding potential risk behaviors. Future work will investigate the general case of involving more urban management departments to achieve better operation efficiency of the entire system.

\newpage

\bibliographystyle{named}
\bibliography{ijcai25}

\end{document}